\documentclass[rnote]{aa}
\usepackage{graphicx}
\usepackage{txfonts}

\newcommand{\mps}{m\,s$^{-1}$}
\newcommand{\bvec}[1]{ \mbox{\boldmath$#1$} }
\def\id{{\rm d}}

\def\cK{{\cal K}}

\begin{document}

   \title{Issues with time--distance inversions for supergranular flows}

   \author{Michal \v{S}vanda
          \inst{1,2}
          }
   \institute{Astronomical Institute, Academy of Sciences of the Czech Republic (v. v. i.), Fri\v{c}ova~298, CZ-25165, Ond\v{r}ejov, Czech Republic\\
              \email{michal@astronomie.cz}
         \and
             Astronomical Institute, Faculty of Mathematics and Physics, Charles University in Prague, V Hole\v{s}ovi\v{c}k\'{a}ch 2, CZ-18000, Prague~8, Czech Republic
             }

   \date{Received 23 October 2014; accepted 26 January 2015}

\abstract{}{Recent studies showed that time--distance inversions for flows start to be dominated by a random noise at a depth of only a few Mm. It was proposed that the ensemble averaging might be a solution to learn about the structure of the convective flows, e.g., about the depth structure of supergranulation.}
{Time--distance inversion is applied to the statistical sample of $\sim 10^4$ supergranules, which allows to regularise weakly about the random-noise term of the inversion cost function and hence to have a much better localisation in space. We compare these inversions at four depths (1.9, 2.9, 4.3, and 6.2~Mm) when using different spatio-temporal filtering schemes in order to gain confidence about these inferences.}
{The flows inferred by using different spatio-temporal filtering schemes are different (even by the sign) even-though the formal averaging kernels and the random-noise levels are very similar. The inverted flows alterates its sign several times with depth. It is suggested that this is due to the inaccuracies in the forward problem that are possibly amplified by the inversion. It is possible that also other time--distance inversions are affected by this issue.}{}

   \maketitle

\section{Motivation}

Helioseismology is considered to be a state-of-the-art method of solar research to learn about the solar interior \citep[see review by][]{2010ARAA..48..289G}. Among various helioseismic methods, the local helioseismology plays an important role in studying the spatially localised details. The time--distance helioseismology \citep{1993Natur.362..430D} is one of the approaches. It consists of tools for measuring and interpretation of travel times of solar waves. Since its formulation, it was used extensively to infer especially information about flows and sound-speed anomalies in upper layers of solar convection zone. 

The method assumes that the travel time $\tau^a$ is bound to the plasma anomalies $\delta q_\alpha$ with respect to the background by a linear equation
\begin{equation}
\delta\tau^a(\bvec{r})=\int_\odot \id^2\bvec{r}' \sum_\beta K^a_\beta(\bvec{r}'-\bvec{r}) \delta q_\beta(\bvec{r}',z)+n^a(\bvec{r}),
\label{eq:forward}
\end{equation}
where $K^a_\alpha$ is the sensitivity kernel coming from the forward modelling, $\bvec{r}=(x,y)$ is a horizontal position vector in a Cartesian coordinate system (the remaining vertical component will be denoted as $z$) and $n^a$ is the travel-time random noise. Index $a$ denotes the selection of the waves (the combination of the spatio-temporal filter, spatial averaging, and the distance between the travel-time measurement points) and greek indices $\alpha$ or $\beta$ select the perturbation $\delta q$ (flow components, sound speed perturbation, \dots). 

One of the main goals of time--distance helioseismology is to infer the structure of convective flows in the near-surface layers of solar convection zone from surface measurements of wave travel times. Formally it means that we replace $\delta q_\beta$ in (\ref{eq:forward}) by $v_\beta$, where $\beta=(x,y,z)$, and search for it. This can only be achieved by inverse modelling. 

To solve the inverse problem, one has to set up the cost function with various terms, which can be cast to a system of linear equations. How to perform such task was in details described and discussed elsewhere. The resulting flow estimate $\mbox{\boldmath$v$}^{\rm inv}$ is given by
\begin{eqnarray}
v^{\rm inv}_\alpha (\bvec{r}_0;z_0) & = & \int_\odot \sum_\beta \cK^\alpha_\beta(\bvec{r}-\bvec{r}_0, z; z_0) v_\beta(\bvec{x}) \; \id^2\bvec{r}\,\id z \nonumber \\
& & + \sum_{i, a} w^\alpha_a(\bvec{r}_i-\bvec{r}_0; z_0) n^a(\bvec{r}_i) ,
\label{eq:inverted}
\end{eqnarray}
where $w^\alpha_a$ are the inversion weights to be determined and $\cK^\alpha_\beta$ indicates the component of the inversion averaging kernel (a linear combination of the sensitivity kernels with inversion weights). The averaging kernel describes the smoothing of the real solar convective flows and is constructed to peak around a target depth $z_0$. The second term on the right-hand side of (\ref{eq:inverted}) indicates the realisation of the random noise. 

For the inversion method which is the base of this study, the reader is referred to \cite{2012SoPh..276...19J} and \cite{Svanda2011}. The cost function contains four terms balanced by three trade-off parameters: the quality of the fit (\emph{the misfit}) of the averaging kernel to a user-defined target function localised in the Sun, the level of the random noise, the level of the pollution of the inverted flow component by other components (\emph{the cross-talk}), and the term which ensures that the resulting inversion weights are spatially localised (which is a requirement in order to fulfil the mathematical assumptions). 

One of the biggest issues in time--distance helioseismology is the presence of the realisation noise. Pressure and surface gravity waves are randomly excited by the vigorous convection and hence their statistics inherently contains a large random component. This acts as a random noise in any of the helioseismic observables and travel times of the waves are no exception. Various helioseismic methods deal with the random noise differently. Sometimes the noise is ignored \citep[e.g.][]{1997ASSL..225..241K}, sometimes the variance of the travel times is considered \citep[e.g.][]{2001ApJ...557..384Z} and in the ideal case, the full travel-time noise covariance matrix is considered \citep[e.g.][]{2005ApJS..158..217C,2008SoPh..251..381J,Svanda2011}. 

The precise knowledge of the noise covariance matrix is essential, as this error propagates through the inversion procedure and translates into the realisation of the random error in the inverted maps. The determination of the full noise-covariance matrix is not straightforward. There are two approaches being used. Either the covariance matrix is read out directly from the data by measuring a large set of travel-time maps, or it is derived from the model using a Monte-Carlo-like approach. Both approaches have their own advantages and disadvantages. The data-driven approach would be probably ideal, however there might be time-varying systematic errors in the travel-time measurements, which will affect also the estimated error level of the inverted quantity. The model-driven methods relies on how precisely does the used model agree with the observed power spectrum of the waves. So it would be ideal if one could minimise the effect imprecise knowledge of travel-time noise to the knowledge of solar flows.

Moreover, several studies \citep{2007ApJ...668.1189W,2008SoPh..251..381J} showed that the inversions for snapshot of the solar flows averaged over a few hours starts to be dominated by the random-noise component already at very shallow depths. 

\begin{figure*}
\centering
\includegraphics[width=0.9\textwidth]{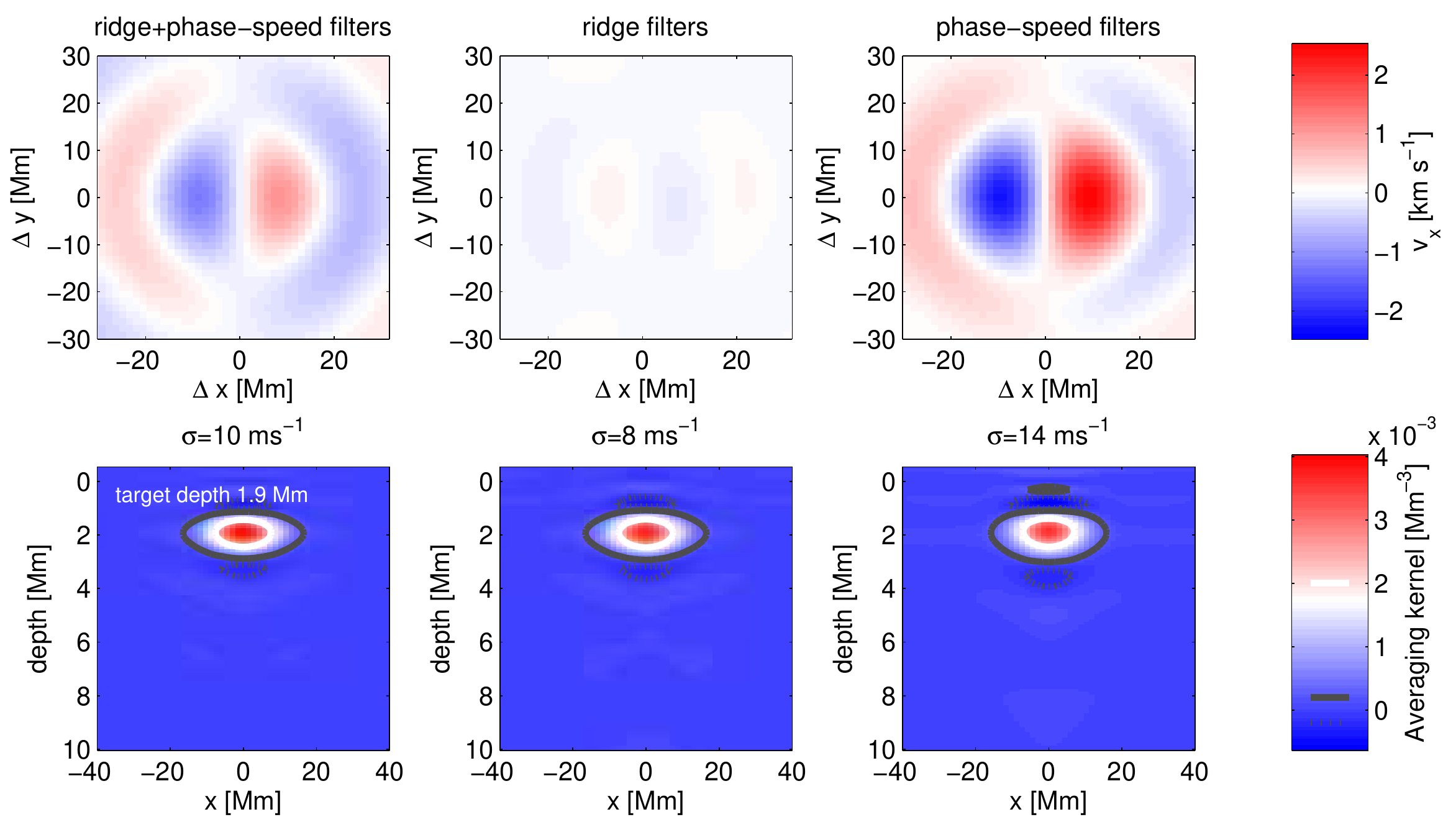}\\
\rule{0.8\textwidth}{1pt}\\
\includegraphics[width=0.9\textwidth]{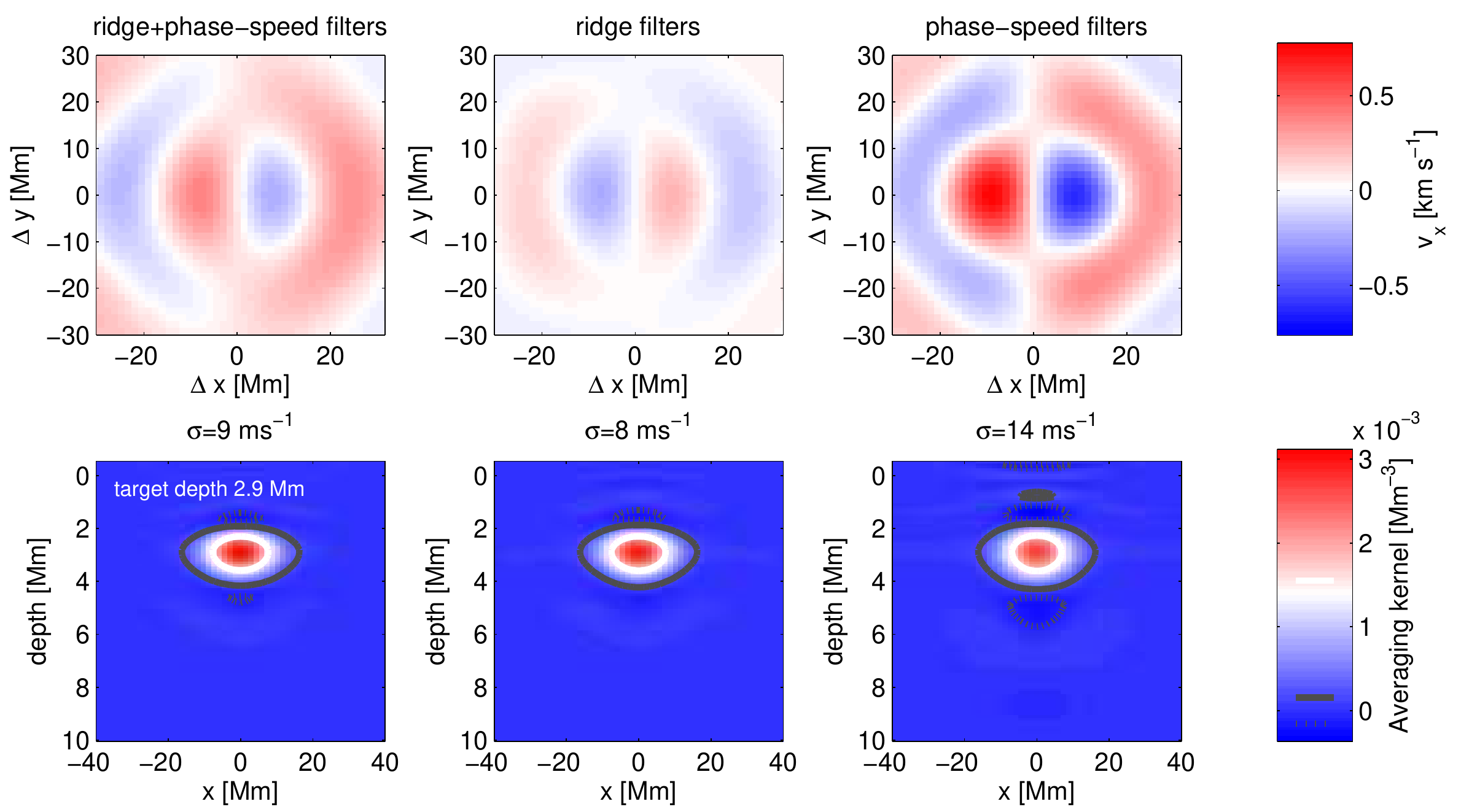}
\caption{The inversions for the horizontal flows in an average supergranule at depths 1.9 and 2.9~Mm, together with the display of the corresponding averaging kernels. The left column represents the inversion utilising combined ridge+phase-speed filtering, the middle column is for the ridge filters only and the right column for phase-speed filters only. }
\label{fig:comparisons}
\end{figure*}

\begin{figure*}
\centering
\includegraphics[width=0.9\textwidth]{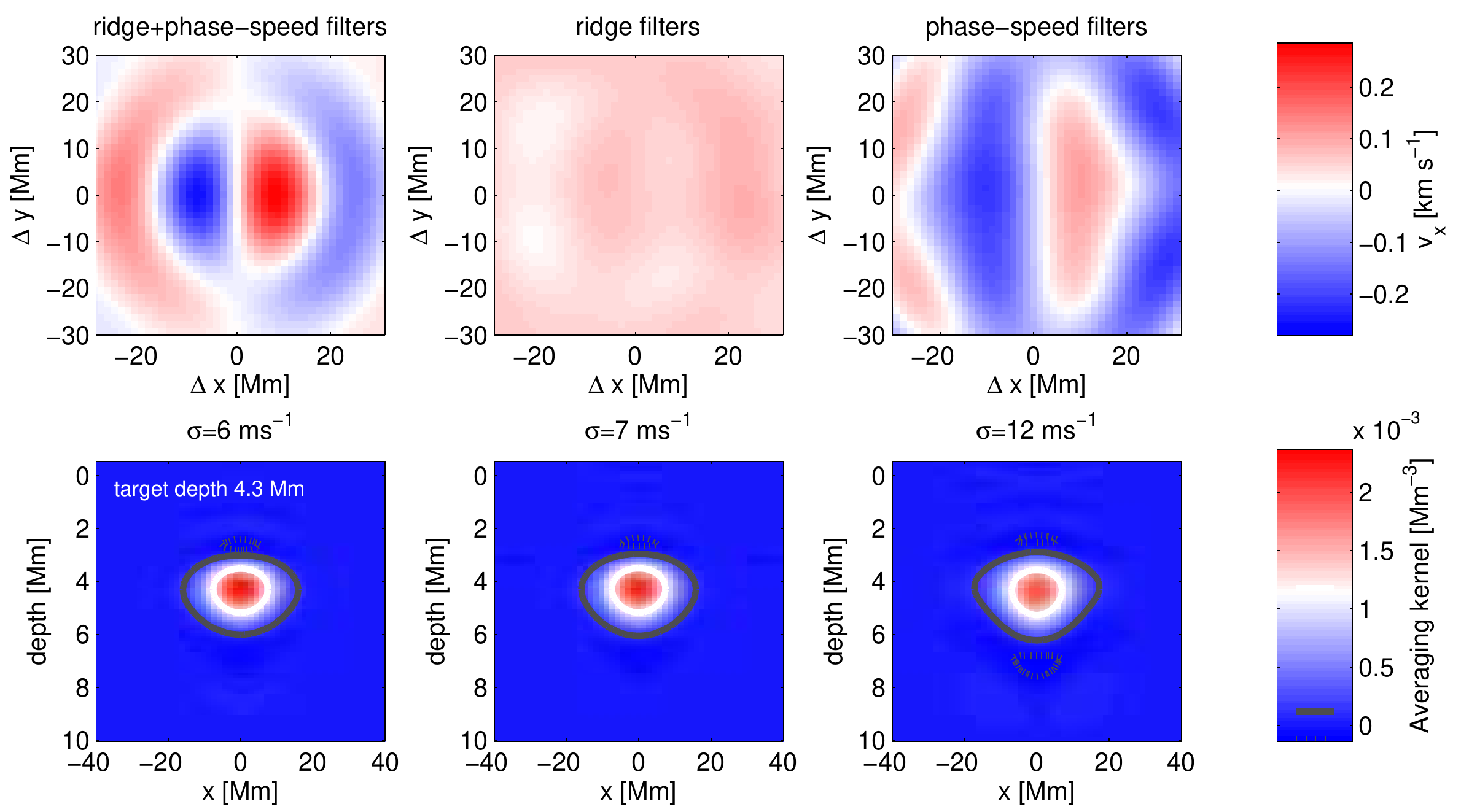}\\
\rule{0.8\textwidth}{1pt} \\
\includegraphics[width=0.9\textwidth]{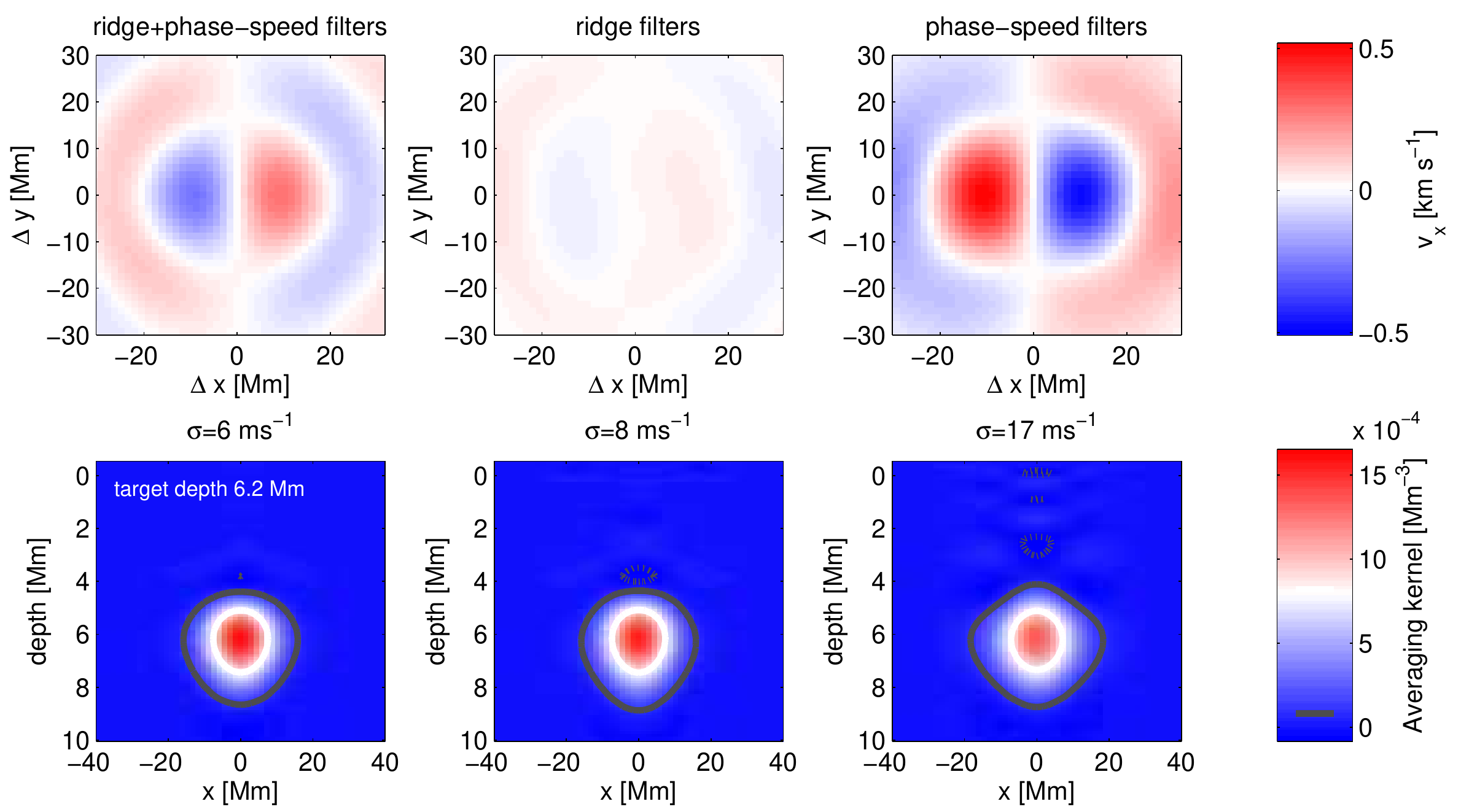}
\caption{Same as Fig.~\ref{fig:comparisons}, but for the depths of 4.3 and 6.2~Mm.}
\label{fig:comparisons2}
\end{figure*}

In \cite{Svanda2011} we suggested the use of the ensemble averaging approach. The ensemble averaging uses a strong assumption that the random-noise realisations in the individual representatives of the ensemble are independent, and hence in the average the random-noise level scales as $1/\sqrt{N}$, where $N$ is the number of representatives. When $N$ is large enough, it allows to relax the noise term in the inversion cost function and thus to decrease the effect of the possible inaccurate knowledge of the noise covariance matrix. The ensemble averaging approach certainly is not useful for studying the snapshots of the flows, however when investigating the set of the representatives of the same phenomenon (supergranules, sunspots,\dots), it seems extremely powerful. In recent years it seems to be a standard method of helioseismic research \citep[e.g.][]{2006ApJ...646..553D,2010ApJ...725L..47D,2012ApJ...759L..29S,2013ApJ...762..131B,2014ApJ...790..135S}. 

\section{Inversion}

The results presented in this note were obtained using the real data coming from the HMI archive. Using the standard tracking and mapping pipeline (created and maintained within the German Science Center for SDO by H.~Schunker and R.~Burston), 24-hours long Dopplergram datacubes were tracked on a daily basis, from 8 May 2010 to 12 July 2010. These datacubes covered the central part of solar disc roughly having 60 degrees on a side in the Postel projection with a cadence of 45~s. Each datacube was processed in a standard way in a travel-time measurement pipeline. From each frame of the datacube, the mean image capturing mostly the pattern of supergranulation was removed. The datacubes then underwent the spatio-temporal filtering for a set of filters ($f$ to $p_4$ ridge filters and eleven standard phase-speed filters) to retain only the waves with desired properties. The travel times were then measured from the filtered datacubes using the linearised \cite{2004ApJ...614..472G} approach for a set of distances for each of the spatio-temporal filters using the centre-to-annulus and centre-to-quadrant averaging schemes. Travel-time maps of the waves sensitive towards the surface (the $f$ mode and first four phase-speed filters) in a centre-to-annulus geometry show clearly the pattern of divergence centres corresponding to supergranulation.

These travel-time maps were inverted in an inversion pipeline. The inversion code is written in {\sc Matlab} language and described in details in \cite{Svanda2011}. It utilises the Born-approximation sensitivity kernels \citep{2007AN....328..228B}\footnote{All point-to-point sensitivity kernels were computed by a code {\sc kc3} written by Aaron Birch, which uses the normal mode summation approach with considered modes upto radial order of 8, quadrupole source at the depth of 100~km, and the observation height of 300~km. The correlation time was chosen to be 48~s. A very fine grid in both the wave-number ($6\times10^{-5}$~Mm$^{-1}$) and frequency (43~$\mu$Hz) space was chosen for a initial computation, however the results do not seem to be extremely sensitive to the grid selection. The point-to-annulus and point-to-quadrant kernels were computed subsequently utilising spatial averaging.} and full travel-time covariance matrix measured directly from a large set of travel-time maps. The inversion implements the Multichannel Subtractive-OLA approach \citep{2012SoPh..276...19J} with additional terms of the cross-talk minimisation and inversion weights localisation. The cost function is regularised strongly about these terms, as they both are possible sources of biases. Both the travel-time and inversion pipelines were validated against the surface measurements \citep{2013ApJ...771...32S}. 

Three different combinations of sensitivity kernels were used. The first set combined all ridge-filtered kernels ($f$ to $p_4$ with a set of annulus radii in the range of 5 to 20 pixels with a step of 1 px, hence 240 independent kernels), second all phase-speed-filtered kernels (eleven standard phase-speed filters with five distances for each of them as tabulated in Table~1 of \cite{2006ApJ...640..516C}, together 165 kernels) and finally the inversion that combined all these kernels at once (405 kernels together). The inversion using the combined filtering scheme was used only recently \citep{2013ApJ...775....7S,2014ApJ...788..127D,2014ApJ...794...18D} and the tests by \cite{2013ApJ...775....7S} showed that for a testing depth they provided flow estimates that were highly comparable. These tests however were performed in a different regime, when the cost function was regularised strongly about the random-noise term. 

A different regime was used in this study by running the inversion suitable for the ensemble averaging approach. In this case, the requirement on the random-noise term of the inversion cost function may be relaxed, which allows the inversion to find the solution with a smaller misfit. In another words, the inversion results in the averaging kernel that is better localised in space and has fewer sidelobes usually making the interpretation difficult.

The testing sample consisted of 17\,474 individual supergranular cells, identified by a watershed algorithm in the centre-to-annulus travel-time maps. The segmentation algorithm and its implementation is described elsewhere \citep{2014ApJ...790..135S}. The resulting flow maps of all three flow components were aligned about the centres of all detected supergranules and averaged, however for the following discussion only the $v_x$ component (the component in the direction of solar rotation) was used. The claims are exactly the same for the remaining horizontal component $v_y$ and the issues to be revealed get even worse for a weak vertical $v_z$ component, for which the random-noise constraint was relaxed beyond the reasonable signal-to-noise ratio. 

\section{Interpretation issues}

The aim of this study was to test the credibility of the time--distance inversions for convective flows. In the misfit-dominated regime the main burden of the inversion quality is carried by the sensitivity kernels which come from forward modelling. Only a slight regularisation is performed by the random-noise term. A strong regularisation about the cross-talk and weights localisation terms should guarantee the minimisation of biases. 

All three different travel-time filtering schemes in the inversion provide similar averaging kernels and similar random-noise levels. Equation (\ref{eq:inverted}) implies that in that case also the estimate of the inverted flow should be similar. Such claim was verified by varying the trade-off parameters for the inversions utilising the same set of measurements. 

\begin{table*}[!t]
\begin{tabular}{l|cccc}
\hline \hline
Inversion & $\rho(v_x^{\rm akern}, v_x^{\rm inv})$ & $\mathrm{RMS}(v_x^{\rm akern})$ & expected noise & derived noise\\
& & [\mps] & [\mps] & [\mps]\\
\hline
1.9 Mm, combined & 0.99 & 70 & 14 & 11 \\
2.9 Mm, combined & 0.99 & 59 & 9 & 8\\
4.3 Mm, combined & 0.99 & 49 & 7 & 6\\
6.2 Mm, combined & 0.99 & 45 & 6 & 5\\
2.9 Mm, ridge & 0.99 & 71 & 14 & 12\\
2.9 Mm, ridge & 0.99 & 59 & 12 & 9\\
4.3 Mm, ridge & 0.99 & 49 & 10 & 8\\
6.2 Mm, ridge & 0.99 & 45 & 8 & 6\\
1.9 Mm, phase-speed & 0.98 & 70 & 16 & 14\\
2.9 Mm, phase-speed & 0.98 & 59 & 15 & 13\\
4.3 Mm, phase-speed & 0.97 & 49 & 15 & 12\\
6.2 Mm, phase-speed & 0.97 & 44 & 15 & 11\\
\hline
\end{tabular}
\caption{Tests of the inversion using the synthetic data. For each inversion, the correlation coeficient between the expected and inverted flow is given, the RMS of the expected flow, the expected RMS of the random-noise component, and its measured value.}
\label{tab:synthetic}
\end{table*}

The inverted flow maps alongside with the corresponding averaging kernels and noise levels are presented in Figs.~\ref{fig:comparisons} and \ref{fig:comparisons2}. It seems that the above mentioned implication does not hold when different sets of measurements are used in inversions. For each investigated depth (1.9, 2.9, 4.3, and 6.2~Mm) one can obtain different answers (even by a sign) just by selection of the different set of measurements, when the random-noise levels are comparable (within the factor of two, given the magnitudes of the inverted flows with very large signal-to-noise ratios) and the averaging kernels are very similar. It has to be noted that due to the strong regularisation of the cost function about the cross-talk term, the components of the averaging kernels that are not in the direction of the inversion (equation (\ref{eq:inverted}), first term on the right-hand side for $\alpha\ne\beta$) are negligible (not displayed). The second apparent issue is that within the inversions based on the same set of measurement, the sign of the flow reverses repeatedly as one goes deeper. This does not seem to be supported from the side of the theory of solar convection. 

The inversions for statistical ensembles are sort of extreme. In this extreme regime we however see clearly that they are not robust. There is no reasons not to believe that in the case when the noise-regularisation term is stronger and the resulting flow maps are seemingly realistic, the problems persist to an unknown (possibly smaller) extent. For instance the number of reversal may be manipulated by the strength of the noise-regularisation term. Generally speaking, the stronger the regularisation, the fewer reversals. It is interesting to point out that in the past the depth of the supergranulation was estimated from the depth where the inverted flow reversed its sign. \cite{1997SoPh..170...63D,1998ESASP.418..581D,2003ESASP.517..417Z,2009NewA...14..429S} reported the reversals depth within a few Mm below the surface. Recently, \cite{2014ApJ...788..127D} used similar time--distance inversion as this study to validate it by using the state-of-the-art simulation of solar convection and saw the supergranular flow reversal, even when it was not present in the simulation. The authors claimed that it was due to the shortcomings of the helioseismic inversion methods. 

\subsection{Verification of the inversion using synthetic data}
A natural explanation of the issues described above would be an error (a programmer's bug) in the inversion code, or a mathematical problem in the inversion itself (e.g., a degeneracy of the matrix to be inverted). To eliminate such possibility, a test involving the synthetic data was applied, similarly to the tests performed by \cite{Svanda2011}.

A snapshot from the simulation of the convective Sun-like flows \citep{2006ASPC..354..115U} was convolved with the appropriate set of the sensitivity kernels, in accordance with Eq.~(\ref{eq:forward}), to create a set of synthetic travel-time maps. A random realisation of the travel-time noise having the covariance matrix that of the inversion was added to these maps to mimic the random excitation of solar waves. In total, three different sets of travel-time maps were computed, one for the ridge filters, one for the phase-speed filters and one for all filters combined. These travel-time maps were inverted by using exactly the same inversion weights as used above. 

The resulting flow maps were compared on the pixel-to-pixel basis to the flow maps obtained by the smoothing of the simulation with the inversion averaging kernels. Essentially, the left-hand side of Eq.~(\ref{eq:inverted}) was directly compared to the first term of the right-hand side of the same equation, denoted as $v_\alpha^{\rm akern}$ henceforth. The root-mean-squared (RMS) value of the random noise term (the second term on the right-hand side of Eq.~(\ref{eq:inverted})) was then compared to the expected level of the random noise, another output of the inversion. The results are summarised in Table.~\ref{tab:synthetic}.

\begin{figure}
\centering
\includegraphics[width=0.4\textwidth]{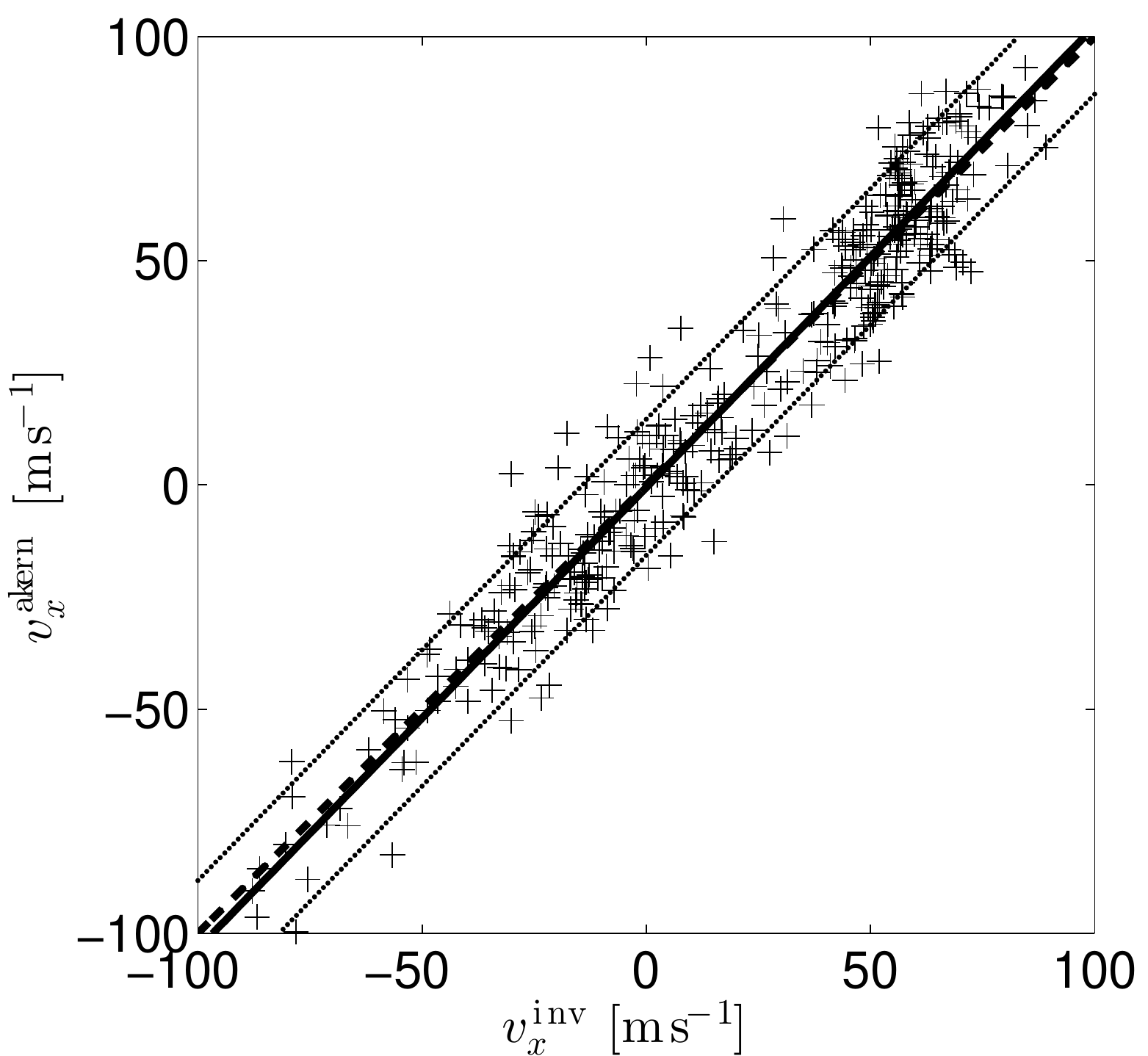}
\caption{A direct comparison of the expected horizontal flow values to the inverted ones on the pixel-to-pixel basis (only each 10th pixel is plotted to make the plot simpler). The solid line represents the linear fit to the crosses, the thin dotted lines the intervals of the predicted random-noise ($1\sigma$ interval), and the thick dashed line is the line with the unity slope.}
\label{fig:crossplots}
\end{figure}

One can see that the inversion behaves as expected. The inverted flow maps $v^{\rm inv}$ are highly correlated to the expected flow maps $v^{\rm akern}$. The determined level of the random noise is consistent with the predicted one from the inversion. The worsts case (i.e. depth 6.2~Mm for the phase-speed kernels, which has the lowest correlation coefficient and the signal-to-noise ratio) is displayed explicitly in Fig.~\ref{fig:crossplots}. Even this worst case scenario demonstrates that the inversion behaves well and no sign of a obvious bug is found. 

\subsection{Does the inverted flow fit the observed travel times?}
\begin{figure*}
\centering
\includegraphics[width=0.9\textwidth]{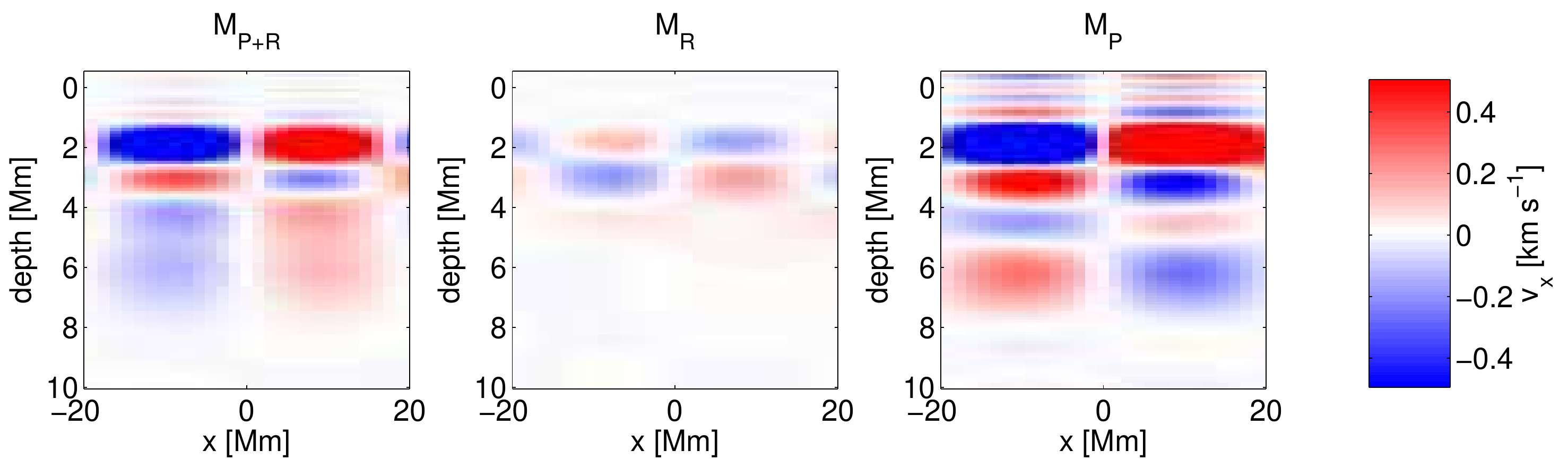}
\caption{Estimated continuous horizontal velocities obtained from four tomographic maps at depths of 1.9, 2.9, 4.3, and 6.4~Mm. The magnitudes are saturated at levels $\pm0.5$~k\mps.}
\label{fig:models}
\end{figure*}

\begin{table}
\centering
\begin{tabular}{l|ccc}
\hline
\hline
& M$_{\rm r+p}$ & M$_{\rm r}$ & M$_{\rm p}$ \\
\hline
$\chi^2_{\rm r+p}$ [s] & 0.24 &   0.24 &  0.32\\
$\chi^2_{\rm r}$ [s] &  0.26 &  0.30 &   0.33\\
$\chi^2_{\rm p}$ [s] &  0.20 &  0.16 & 0.32 \\
\hline\hline
\end{tabular}
\caption{Match of the forward-modelled travel times for three considered models to the observed ones.}
\label{tab:fits}
\end{table}

In SOLA inversions it is not guaranteed that the inverted flow model provides a reasonable fit to the observed travel times. It is obvious already from Eq.~(\ref{eq:forward}): in order to obtain forward-modelled travel times, the knowledge of the continuous flow model is needed, which is not returned by SOLA. A proper reconstruction is probably possible (a work in progress), however for a simple illustration a much rougher estimate $v_\alpha^{\rm est}$ can be made: 
\begin{equation}
v_\alpha^{\rm est}(\bvec{r},z)=\sum\limits_{z_0} v_\alpha^{\rm inv} (\bvec{r};z_0) \cK_\alpha^{\rm 1D}(z;z_0),
\end{equation}
where $\cK_\alpha^{\rm 1D}=\int\id^2\bvec{r}\,\cK_\alpha^\alpha(\bvec{r},z;z_0)$ is a horizontally averaged averaging kernel. Such an estimate approximately fulfills Eq.~(\ref{eq:inverted}) with the random-noise term excluded. The $v_x^{\rm est}$ components of three different models from inversions based on different filtering schemes are shown in Fig.~\ref{fig:models} -- $M_{\rm P+R}$ for the combined ridge+phase-speed filtering scheme, $M_{\rm R}$ for ridge and $M_{\rm P}$ for phase-speed filters based inversions respectively. The suspicious alterations of the sign of the horizontal flow with depth are prominently visible. Models $M_{\rm P+R}$ and $M_{\rm P}$ are similar at depths less than 5~Mm and differ by sign deeper down. Model $M_{\rm R}$ is significantly different from the other two. 

From these models, forward-modelled travel-time maps were computed following Eq.~(\ref{eq:forward}) with the noise term neglected. Such neglection was fully justified by the ensemble averaging technique, which increased the signal-to-noise of the measured travel-time maps by more than two orders. A match to the observed travel times $\tau^a$ to the forward-modelled ones $\tau^a_{\rm fm}$ was estimated from
\begin{equation}
\chi^2_{\rm fm}=\frac1M \sum\limits_a {\rm RMS}\,\left( \tau^a_{\rm fm}(\bvec{r}) - \tau^a (\bvec{r}) \right),
\end{equation}
where $M$ was the total number of travel-time measurements indexed by $a$. The larger $\chi^2_{\rm fm}$, the worse the forward-modelled travel times fit the observed ones. 

The fit was evaluated for each considered vector-flow model for all three sets of travel-time measurements separately. The results are summarised in Table~\ref{tab:fits}. One would naively expect the main-diagonal terms be much smaller than the off-diagonal terms. From a visual inspection both the observed and forward-modelled travel-time maps it becomes clear that all the fits are far from being satisfactory. Dispite being very different (even by a sign), all three models fit the observed travel times equally (badly). 

The simple reconstruction of the continuous flow field probably also affects the results, but given the fact that the alteration of the sign is visible already in the tomographic maps, the averaging kernels are very confined around the target depth and have negligible sidelobes, it can be expected that a more realistic estimate of the continuous flow field will not dramatically change the conclusions. 

\section{Possible causes}
\begin{figure*}
\includegraphics[width=0.49\textwidth]{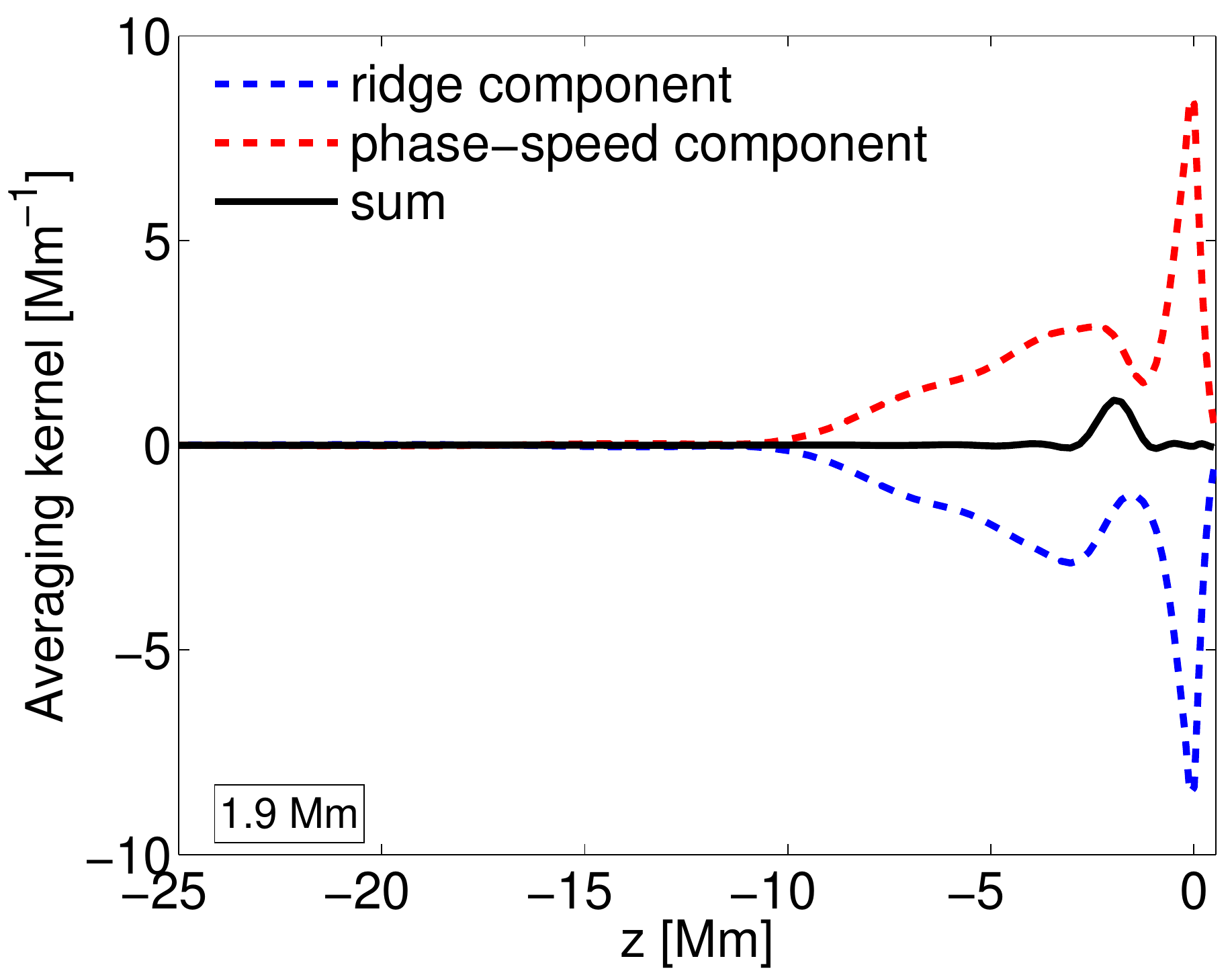}
\includegraphics[width=0.49\textwidth]{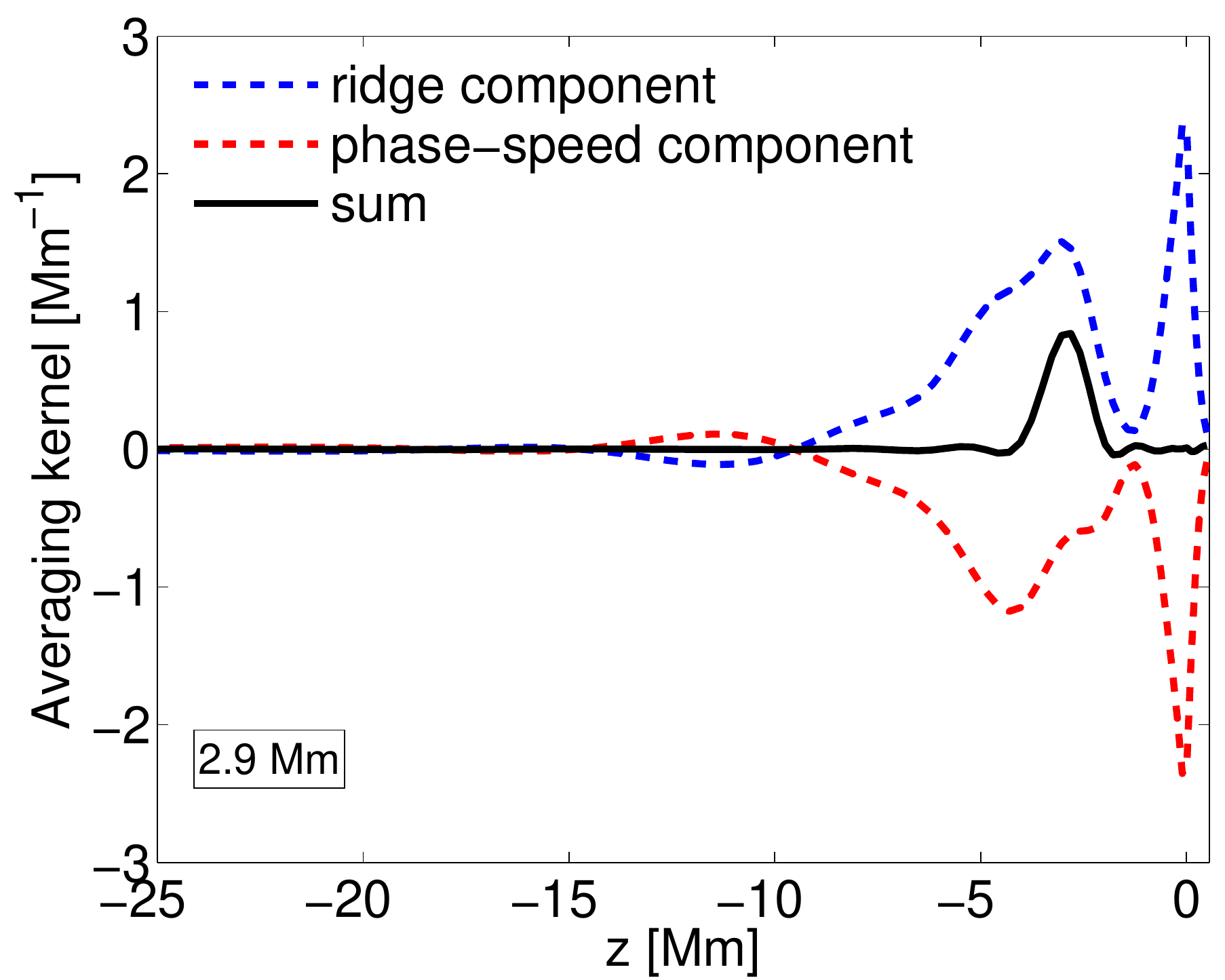}\\
\includegraphics[width=0.49\textwidth]{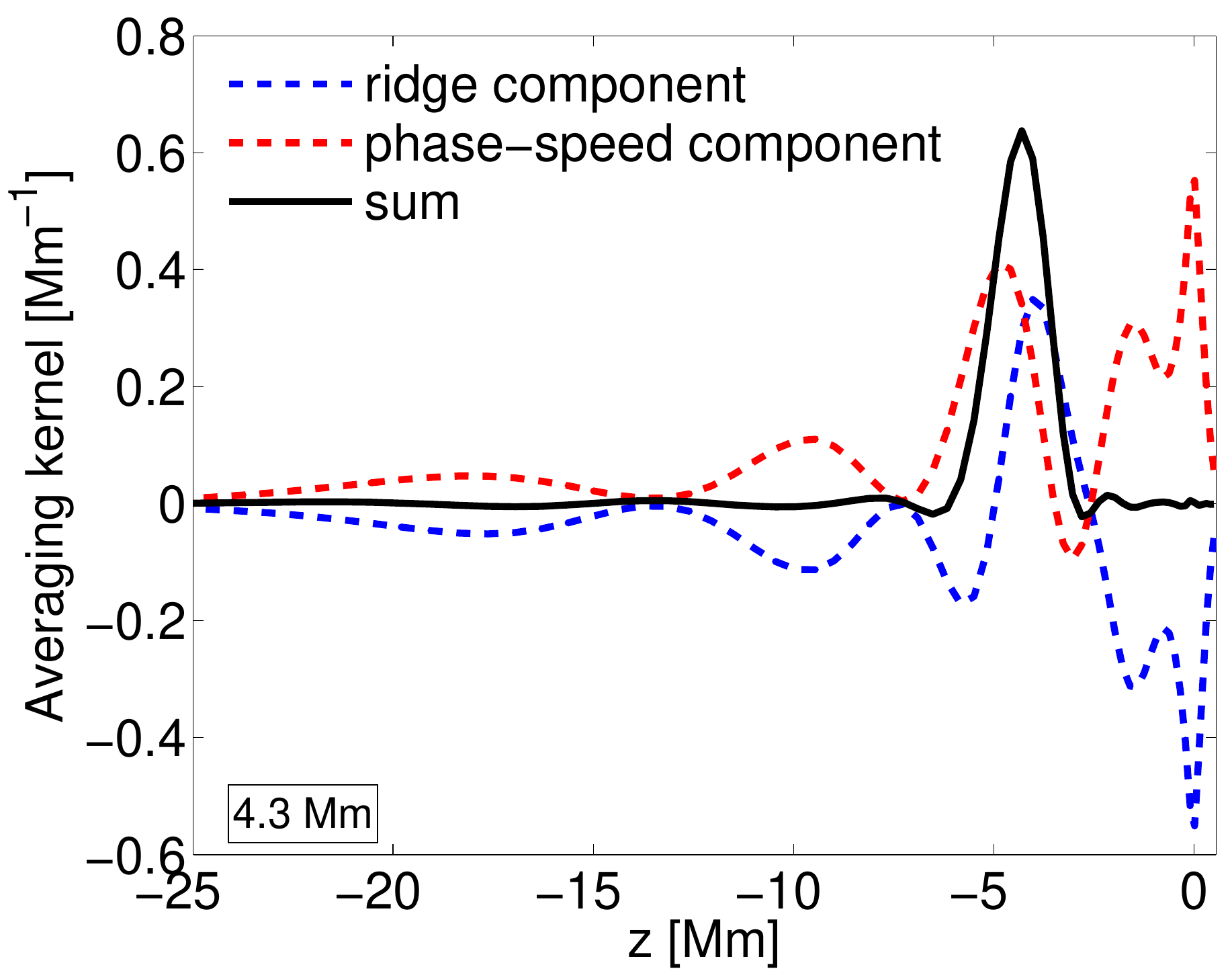}
\includegraphics[width=0.49\textwidth]{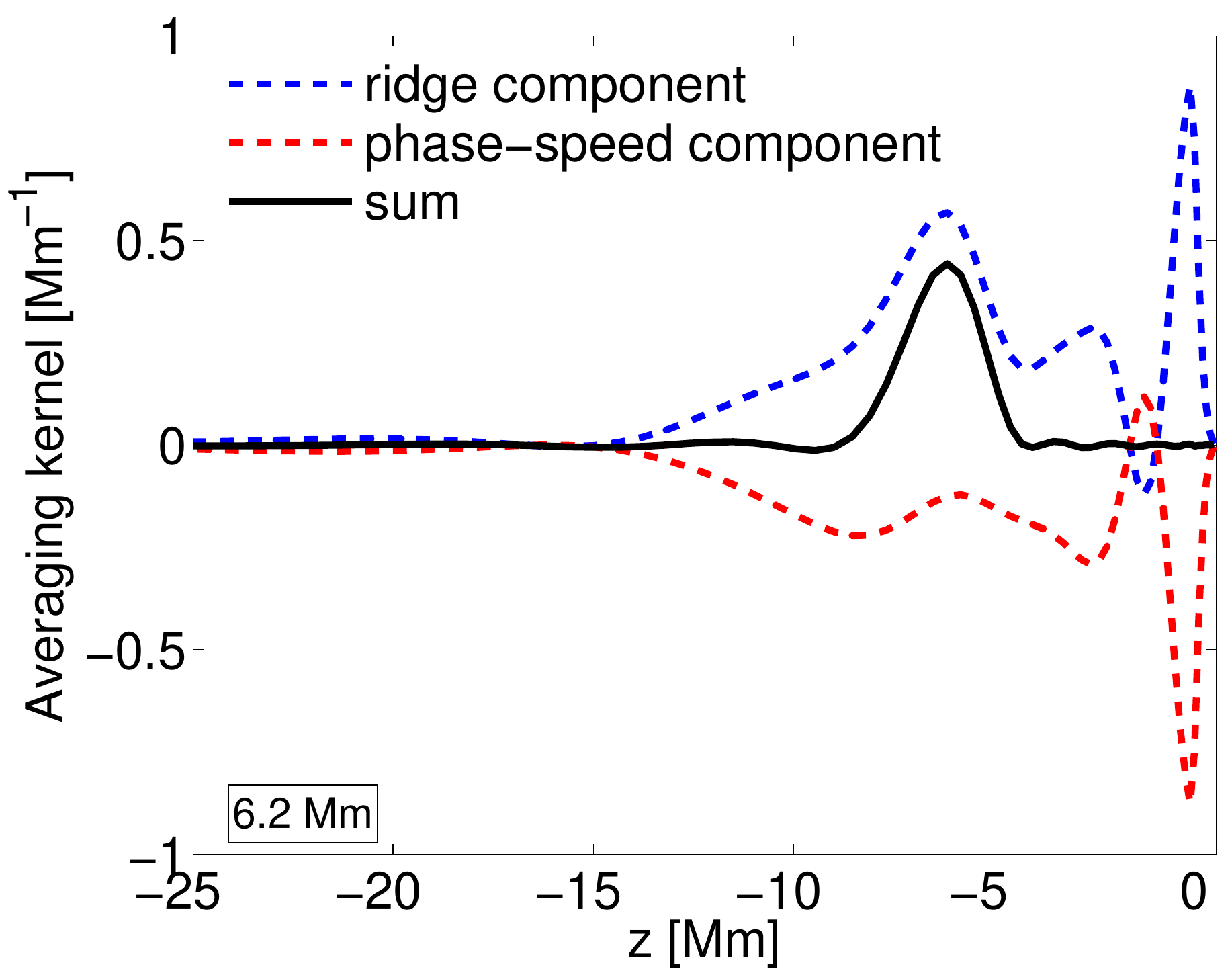}
\caption{The contributions of various filtering schemes to the inversion averaging kernel in case of the combined filtering scheme.}
\label{fig:akerncomps}
\end{figure*}

There are several suspicious contributors possibly responsible for the issues. 

{\bf Incompatibility of the travel time maps with the inversion weights.} As it was pointed out many times in the past by various authors, it is crucial that any step in the data processing (mapping, filtering, the way the travel times are measured) must be taken into account when computing the corresponding sensitivity kernels. In our case, this need was enforced as much as possible. The pixel size was exactly the same in case of both the data processing and the computation sensitivity kernels, the spatio-temporal filtering was done not only using the same code, but moreover using the exactly same files, in which the filters were stored. The Born sensitivity kernels \citep{2007AN....328..228B} are consistent with the \cite{2004ApJ...614..472G} definition of the travel times. The issue with unknown impact are the possible non-linearities in both the forward and inverse problems. 

{\bf Difference of the power spectrum of the real data and the model.} The sensitivity kernels are computed using the power spectrum, which comes from the model, which is considered Sun-like \citep{1996Sci...272.1286C}. However, the eigenfrequencies deviate from the data eigenfrequencies for higher order modes (already a $p_6$ mode frequencies are considerably off), the eigenfrequencies are not available for very small wave numbers ($k<0.1$~Mm$^{-1}$). To investigate the influence of this issue, I recomputed sensitivity kernels, travel times, and inversions with an additional spatio-temporal filter, which limited the power spectrum to the region where the model power spectrum matched well to the power spectrum of the used datacubes. Hence low wave numbers were filtered out, as was all the signal of oscillations beyond the $p_6$ ridge. The introduction of this additional filter had only minor impact on the results. 

{\bf Inaccuracy of sensitivity kernels.} Should the sensitivity kernels forward-modelled from a reference solar model be different from ``those of the real Sun'', we must expect an impact of this deviation to the results. The difference is expected to be small if the reference model is Sun-like. Unfortunately, there does not seem to be a direct way to measure the sensitivity kernels from solar data, perhaps except for the iterative inversions \citep{2014ApJ...784...69H,2014arXiv1410.1981H}, introduced to helioseismology only recently. There might be the way to at least verify the total integral of the kernel, which is the work in progress and will be reported on in a later paper. This point embodies both the effects of the differences between the reference model and the real Sun and the possible issues with the kernel computation, both theoretical and numerical. This point is also supported by \cite{2014ApJ...788..127D}, where the authors using a realistic convection simulation noticed a significantly decreasing correlation between the measured and forward-modelled travel times for ridge filters with higher orders ($p_3$ and beyond) and also for larger phase speeds. 

{\bf Mathematically posed problem.} This possibility is closely related to the previous point. Let's say that the sensitivity kernels are accurate (again, by ``accurate'' it is ment ``those of the real Sun'') to within 1\%. The combined inversion combines 405 such kernels. These 1\% errors translate through the inversion and may be amplified to an unknown extent. There are some hints that this is likely the most probably cause. The contributions of the averaging kernel from two different filtering schemes are displayed in Fig.~\ref{fig:akerncomps}. One can see that in all four cases discussed in this research note the two components from different filtering schemes largely subtract from each other. Should one filtering scheme have an unknown systematical bias, the resulting ``real'' averaging kernel (i.e. the averaging kernels obtained by convolving inversion weights with the sensitivity kernels ``of the real Sun'', which are unknown) is different and that would easily explain even the change of sign of the flow. A curious reader may object that the way out is not to combine the two filtering schemes. It is shown in Fig.~\ref{fig:comparisons} that the inverted horizontal flow based solely on the phase-speed filtering scheme reverses its sign between the depths of 1.9~Mm and 2.9~Mm, which is not realistic. One has to bear in mind that the subtractions occur in every inversion, e.g. the contribution of point-to-annulus measurements is subtracted from the contribution of the point-to-quadrant measurements etc. Another indication for this explanation is that the magnitude of the inverted flow which uses the phase-speed filters seems to be a bit unrealistic (around 2~k\mps{} for the depth of 1.9~Mm).

Both last points implicate that the ``real'' inversion averaging kernels may be very different from those predicted by the inversion, perhaps even having a negative total integral. Then also the cross-talk contribution is not constrained, which is estimated from the cross-talk components of the averaging kernel. 

\section{Lessons learned}
This work is based on many tens of thousands of CPU-hours trial-and-error runs. In case of the inversions suitable for the tomography of the flow snapshot \citep[see e.g.][]{2013ApJ...775....7S} the issues are not prominently visible, however the question is, whether the stronger regularisation of the solution (about the random-noise term in this case) removes the issues or rather hides them. By studying the literature on flow inversions in supergranules one has to conclude that all the inversions except for the inversion of \cite{2007ApJ...668.1189W} indicated the reversal of the horizontal flow at various depths. It may be seen suspicious, because the state-of-the-art simulations \citep{2009ASPC..416..421S,2009ApJ...691..640R} do not indicate such reversals. It may also easily be that the large-amplitude flows in supergranulation recently reported by time--distance inversions \citep{2012ApJ...759L..29S} are artefacts of similar issues, as in that case $f$ mode and $p_1$ and $p_2$ ridges were used, necessarily leading to subtractions in the inversion.  

From comparisons of the inversion results with the direct surface measurements \citep[e.g.][]{2005ASPC..346....3A,2007ApJ...657.1157G,2007SoPh..241...27S,2013ApJ...771...32S} it seems that the very near-surface inversions (hence involving $f$ modes or acoustic waves with small phase speeds) do not suffer from the discussed problems. As we showed recently \citep{2013ApJ...771...32S}, the flow inversion using the $f$ mode ridge is not only highly correlated with the inferences from the surface granule tracking, but it also provides the properly scaled magnitude of the flow. Such validations against the independently obtained measurements fully justify the scientific results obtained on surface flow fields. It has to be pointed out that no larger subtractions occured in those inversions.

\begin{acknowledgements}
This work was supported by the Czech Science Foundation (grant P209/12/P568). All computations were performed using the {\sc Sunquake} compute cluster at Astronomical Institute of Academy of Sciences in Ond\v{r}ejov, the tracked and mapped datacubes were obtained from data processing pipelines at Max-Planck-Institut f\"ur Sonnensystemforschung (MPS), G\"ottingen, Germany, which is funded by the German Aerospace Center (DLR). The solar measurements were kindly provided by the HMI consortium. The point-to-point travel-time sensitivity kernels were obtained using the code of Aaron Birch deployed at the MPS. This work was done within the institute research project RVO:67985815 to Astronomical Institute of Czech Academy of Sciences. 

This work benefits from couloir discussions following the talk \emph{\v{S}vanda: Current issues with time--distance inversions for flows at As\'U} at \emph{LWS Helioseismology Workshop \#4: Solar Subsurface Flows from Helioseismology: Problems and Prospects} which was held at Stanford University, July 21--23, 2014. 
\end{acknowledgements}


\begin{thebibliography}{34}
\expandafter\ifx\csname natexlab\endcsname\relax\def\natexlab#1{#1}\fi

\bibitem[{{Ambro{\v z}}(2005)}]{2005ASPC..346....3A}
{Ambro{\v z}}, P. 2005, in Astronomical Society of the Pacific Conference
  Series, Vol. 346, Large-scale Structures and their Role in Solar Activity,
  ed. K.~{Sankarasubramanian}, M.~{Penn}, \& A.~{Pevtsov}, 3

\bibitem[{{Birch} {et~al.}(2013){Birch}, {Braun}, {Leka}, {Barnes}, \&
  {Javornik}}]{2013ApJ...762..131B}
{Birch}, A.~C., {Braun}, D.~C., {Leka}, K.~D., {Barnes}, G., \& {Javornik}, B.
  2013, \apj, 762, 131

\bibitem[{{Birch} \& {Gizon}(2007)}]{2007AN....328..228B}
{Birch}, A.~C. \& {Gizon}, L. 2007, Astron.~Nachr., 328, 228

\bibitem[{{Christensen-Dalsgaard} {et~al.}(1996){Christensen-Dalsgaard},
  {Dappen}, {Ajukov}, {Anderson}, {Antia}, {Basu}, {Baturin}, {Berthomieu},
  {Chaboyer}, {Chitre}, {Cox}, {Demarque}, {Donatowicz}, {Dziembowski},
  {Gabriel}, {Gough}, {Guenther}, {Guzik}, {Harvey}, {Hill}, {Houdek},
  {Iglesias}, {Kosovichev}, {Leibacher}, {Morel}, {Proffitt}, {Provost},
  {Reiter}, {Rhodes}, {Rogers}, {Roxburgh}, {Thompson}, \&
  {Ulrich}}]{1996Sci...272.1286C}
{Christensen-Dalsgaard}, J., {Dappen}, W., {Ajukov}, S.~V., {et~al.} 1996,
  Science, 272, 1286

\bibitem[{{Couvidat} {et~al.}(2006){Couvidat}, {Birch}, \&
  {Kosovichev}}]{2006ApJ...640..516C}
{Couvidat}, S., {Birch}, A.~C., \& {Kosovichev}, A.~G. 2006, \apj, 640, 516

\bibitem[{{Couvidat} {et~al.}(2005){Couvidat}, {Gizon}, {Birch}, {Larsen}, \&
  {Kosovichev}}]{2005ApJS..158..217C}
{Couvidat}, S., {Gizon}, L., {Birch}, A.~C., {Larsen}, R.~M., \& {Kosovichev},
  A.~G. 2005, \apjs, 158, 217

\bibitem[{{DeGrave} {et~al.}(2014{\natexlab{a}}){DeGrave}, {Jackiewicz}, \&
  {Rempel}}]{2014ApJ...794...18D}
{DeGrave}, K., {Jackiewicz}, J., \& {Rempel}, M. 2014{\natexlab{a}}, \apj, 794,
  18

\bibitem[{{DeGrave} {et~al.}(2014{\natexlab{b}}){DeGrave}, {Jackiewicz}, \&
  {Rempel}}]{2014ApJ...788..127D}
{DeGrave}, K., {Jackiewicz}, J., \& {Rempel}, M. 2014{\natexlab{b}}, \apj, 788,
  127

\bibitem[{{Duvall}(1998)}]{1998ESASP.418..581D}
{Duvall}, Jr., T.~L. 1998, in ESA Special Publication, Vol. 418, Structure and
  Dynamics of the Interior of the Sun and Sun-like Stars, ed. S.~{Korzennik},
  581

\bibitem[{{Duvall} \& {Birch}(2010)}]{2010ApJ...725L..47D}
{Duvall}, Jr., T.~L. \& {Birch}, A.~C. 2010, \apjl, 725, L47

\bibitem[{{Duvall} {et~al.}(2006){Duvall}, {Birch}, \&
  {Gizon}}]{2006ApJ...646..553D}
{Duvall}, Jr., T.~L., {Birch}, A.~C., \& {Gizon}, L. 2006, \apj, 646, 553

\bibitem[{{Duvall} {et~al.}(1993){Duvall}, {Jefferies}, {Harvey}, \&
  {Pomerantz}}]{1993Natur.362..430D}
{Duvall}, Jr., T.~L., {Jefferies}, S.~M., {Harvey}, J.~W., \& {Pomerantz},
  M.~A. 1993, Nature, 362, 430

\bibitem[{{Duvall} {et~al.}(1997){Duvall}, {Kosovichev}, {Scherrer}, {Bogart},
  {Bush}, {de Forest}, {Hoeksema}, {Schou}, {Saba}, {Tarbell}, {Title},
  {Wolfson}, \& {Milford}}]{1997SoPh..170...63D}
{Duvall}, Jr., T.~L., {Kosovichev}, A.~G., {Scherrer}, P.~H., {et~al.} 1997,
  \solphys, 170, 63

\bibitem[{{Georgobiani} {et~al.}(2007){Georgobiani}, {Zhao}, {Kosovichev},
  {Benson}, {Stein}, \& {Nordlund}}]{2007ApJ...657.1157G}
{Georgobiani}, D., {Zhao}, J., {Kosovichev}, A.~G., {et~al.} 2007, \apj, 657,
  1157

\bibitem[{{Gizon} \& {Birch}(2004)}]{2004ApJ...614..472G}
{Gizon}, L. \& {Birch}, A.~C. 2004, \apj, 614, 472

\bibitem[{{Gizon} {et~al.}(2010){Gizon}, {Birch}, \&
  {Spruit}}]{2010ARAA..48..289G}
{Gizon}, L., {Birch}, A.~C., \& {Spruit}, H.~C. 2010, \araa, 48, 289

\bibitem[{{Hanasoge}(2014)}]{2014arXiv1410.1981H}
{Hanasoge}, S.~M. 2014, ArXiv e-prints

\bibitem[{{Hanasoge} \& {Tromp}(2014)}]{2014ApJ...784...69H}
{Hanasoge}, S.~M. \& {Tromp}, J. 2014, \apj, 784, 69

\bibitem[{{Jackiewicz} {et~al.}(2012){Jackiewicz}, {Birch}, {Gizon},
  {Hanasoge}, {Hohage}, {Ruffio}, \& {\v{S}vanda}}]{2012SoPh..276...19J}
{Jackiewicz}, J., {Birch}, A.~C., {Gizon}, L., {et~al.} 2012, \solphys, 276, 19

\bibitem[{{Jackiewicz} {et~al.}(2008){Jackiewicz}, {Gizon}, \&
  {Birch}}]{2008SoPh..251..381J}
{Jackiewicz}, J., {Gizon}, L., \& {Birch}, A.~C. 2008, Sol. Phys., 251, 381

\bibitem[{{Kosovichev} \& {Duvall}(1997)}]{1997ASSL..225..241K}
{Kosovichev}, A.~G. \& {Duvall}, Jr., T.~L. 1997, in Astrophysics and Space
  Science Library, Vol. 225, SCORe'96 : Solar Convection and Oscillations and
  their Relationship, ed. {F.~P.~Pijpers, J.~Christensen-Dalsgaard, \&
  C.~S.~Rosenthal}, 241--260

\bibitem[{{Rempel} {et~al.}(2009){Rempel}, {Sch{\"u}ssler}, \&
  {Kn{\"o}lker}}]{2009ApJ...691..640R}
{Rempel}, M., {Sch{\"u}ssler}, M., \& {Kn{\"o}lker}, M. 2009, \apj, 691, 640

\bibitem[{{Stein} {et~al.}(2009){Stein}, {Nordlund}, {Georgoviani}, {Benson},
  \& {Schaffenberger}}]{2009ASPC..416..421S}
{Stein}, R.~F., {Nordlund}, {\AA}., {Georgoviani}, D., {Benson}, D., \&
  {Schaffenberger}, W. 2009, in Astronomical Society of the Pacific Conference
  Series, Vol. 416, Solar-Stellar Dynamos as Revealed by Helio- and
  Asteroseismology: GONG 2008/SOHO 21, ed. M.~{Dikpati}, T.~{Arentoft},
  I.~{Gonz{\'a}lez Hern{\'a}ndez}, C.~{Lindsey}, \& F.~{Hill}, 421

\bibitem[{{\v{S}vanda}(2012)}]{2012ApJ...759L..29S}
{\v{S}vanda}, M. 2012, \apjl, 759, L29

\bibitem[{{\v{S}vanda} {et~al.}(2011){\v{S}vanda}, {Gizon}, {Hanasoge}, \&
  {Ustyugov}}]{Svanda2011}
{\v{S}vanda}, M., {Gizon}, L., {Hanasoge}, S.~M., \& {Ustyugov}, S.~D. 2011,
  \aap, 530, A148

\bibitem[{{\v{S}vanda} {et~al.}(2009){\v{S}vanda}, {Klva{\v n}a}, {Sobotka},
  {Kosovichev}, \& {Duvall}}]{2009NewA...14..429S}
{\v{S}vanda}, M., {Klva{\v n}a}, M., {Sobotka}, M., {Kosovichev}, A.~G., \&
  {Duvall}, T.~L. 2009, New Astron., 14, 429

\bibitem[{{\v{S}vanda} {et~al.}(2007){\v{S}vanda}, {Zhao}, \&
  {Kosovichev}}]{2007SoPh..241...27S}
{\v{S}vanda}, M., {Zhao}, J., \& {Kosovichev}, A.~G. 2007, \solphys, 241, 27

\bibitem[{{Ustyugov}(2006)}]{2006ASPC..354..115U}
{Ustyugov}, S.~D. 2006, in Astronomical Society of the Pacific Conference
  Series, Vol. 354, Solar MHD Theory and Observations: A High Spatial
  Resolution Perspective, ed. {J.~Leibacher, R.~F.~Stein, \& H.~Uitenbroek},
  115

\bibitem[{{{\v S}vanda}(2013)}]{2013ApJ...775....7S}
{{\v S}vanda}, M. 2013, \apj, 775, 7

\bibitem[{{{\v S}vanda} {et~al.}(2013){{\v S}vanda}, {Roudier}, {Rieutord},
  {Burston}, \& {Gizon}}]{2013ApJ...771...32S}
{{\v S}vanda}, M., {Roudier}, T., {Rieutord}, M., {Burston}, R., \& {Gizon}, L.
  2013, \apj, 771, 32

\bibitem[{{{\v S}vanda} {et~al.}(2014){{\v S}vanda}, {Sobotka}, \&
  {B{\'a}rta}}]{2014ApJ...790..135S}
{{\v S}vanda}, M., {Sobotka}, M., \& {B{\'a}rta}, T. 2014, \apj, 790, 135

\bibitem[{{Woodard}(2007)}]{2007ApJ...668.1189W}
{Woodard}, M.~F. 2007, \apj, 668, 1189

\bibitem[{{Zhao} \& {Kosovichev}(2003)}]{2003ESASP.517..417Z}
{Zhao}, J. \& {Kosovichev}, A.~G. 2003, in ESA Special Publication, Vol. 517,
  GONG+ 2002. Local and Global Helioseismology: the Present and Future, ed.
  H.~{Sawaya-Lacoste}, 417--420

\bibitem[{{Zhao} {et~al.}(2001){Zhao}, {Kosovichev}, \&
  {Duvall}}]{2001ApJ...557..384Z}
{Zhao}, J., {Kosovichev}, A.~G., \& {Duvall}, Jr., T.~L. 2001, \apj, 557, 384

\end{thebibliography}
\end{document}